# Chapter 1
# Overview of Complex System Design

*John W. Sheppard*

This book is focused on the broad issues surrounding "realizing complex integrated systems." What do we mean by this? It might be helpful to break down our title, one word at a time. To do this, we start at the end. What is a "system"? We start this introductory chapter by addressing that question. We pose a number of possible definitions and perspectives, but leave open the opportunity to consider the system from the target context where it will be used.

Once we have a system in mind, we acknowledge the fact that this system needs to "integrate" a variety of pieces, components, subsystems, in order for it to accomplish its task. Therefore, we concern ourselves at the boundaries and interfaces of different technologies and disciplines in order to determine how best to achieve that integration.

Next we raise the specter that this integrated system is going to be "complex." Complexity can be defined in a number of ways. For one, the sheer number of subsystems or components can be a measure of complexity. Alternatively, we could consider the functions being performed by the system and how those functions interact with one another. Alternatively, we could consider computational aspects such as the amount of time or the amount of memory that may be needed to accomplish one or more tasks. The extent to which new behaviors might emerge from the system can also be regarded as an element of complexity. In the end, complexity is that characteristic of a system that defines the associated challenges along the life of the system, so in this book, we are concerned with how to go about managing that complexity.

Finally, we suggest that "realization" refers to the process by which our complex integrated system moves from concept to deployment and subsequent support. It refers to the entire design, development, manufacture, deployment, operation, and support life cycle. Of particular interested here, however, is that we focus on systems that, by their very nature, are "complex." In other words, we are interested in large, complicated, interacting beasts that are intended to perform difficult tasks and meet a wide variety of end-user needs.

## 1.1 Complex Systems

### 1.1.1 Systems Definition

To begin our discussion of realizing the design of complex systems, we first must talk about what constitutes a system. There are a variety of definitions one could use, and to be sure, many of these definitions will apply here. The natural place to begin is with the dictionary. Here are two dictionary definitions:

> **Definition 1.** (Oxford English Dictionary) A system is a set of things working together as parts of a mechanism or an interconnecting network; a complex whole (OED, 2018).
>
> **Definition 2.** (Merriam-Webster Dictionary) A system is a regularly interacting or interdependent group of items forming a unified whole (Merriam, 2018).

A unifying theme in these two definitions is the notion of parts working together as a whole. Given this, we also consider definitions drawn from different sources focusing on systems engineering. Here are three.

**Definition 3.** A system is the combination of elements that function together to produce the capability required to meet a need. The elements include all hardware, software, equipment, facilities, personnel, processes, and procedures needed for this purpose; that is, all things required to produce system-level results. The results include system-level qualities, properties, characteristics, functions, behavior, and performance (NASA, 2007).

**Definition 4.** A system is an open set of complementary interacting parts, with properties, capabilities, and behaviors emerging, both from the parts and from their interactions, to synthesize a unified whole (Hitchins, 2007).

**Definition 5.** A system is a set of interrelated components working together toward some common objective (Kossiakoff et al., 2011).

The theme persists; however, we are now beginning to see the idea emerge that a system accomplishes some function. If it fails to accomplish that function, then the system does not meet its purpose and is not useful. From an engineering perspective, this translates into the system failing to satisfy its requirements.

From the perspective of one of the editors of this work, we also need to consider what we mean by "system" from the perspective of managing that system's health. Thus we need to consider how system health can be incorporated into the design process. To do so, we acknowledge that one of the added performance requirements for a system is that it needs to be able to be monitored in such a way that its health can actually be determined. However, a related issue in systems engineering arises from the fact the systems, due to their interacting "qualities, properties, characteristics, functions, behavior, and performance" involve complexities that can make such monitoring problematic. Using this observation, Sheppard and Simpson define a system in the context of system diagnostics as follows.

**Definition 6.** A system is any aggregation of related elements that together form an entity of sufficient complexity for which it is impractical to treat all of the elements at the lowest level of detail (Sheppard and Simpson, 1998).

One of the key drivers in any design activity is managing complexity. As the Sheppard and Simpson definition suggests, a system by its very nature has a level of complexity that makes full-system analysis generally impractical. To support the design, analysis, and support of systems, therefore, we need to employ carefully crafted design practices that draw on a variety of domains and disciplines. As we employ these practices, the goal of this book is to draw on the various domains and disciplines with system health always before us.

## 1.1.2 Systems Domains

In a large number of engineering disciplines, system design is presented in a vacuum. Perhaps the three biggest exceptions would be Industrial Engineering, Software Engineering, and Systems Engineering. For our purposes, we approach the engineering problem from the Systems Engineering perspective. Doing so, we recognize and affirm that successful system design must draw from a variety of domains, including the specific technical domain(s) within with the system is being developed.

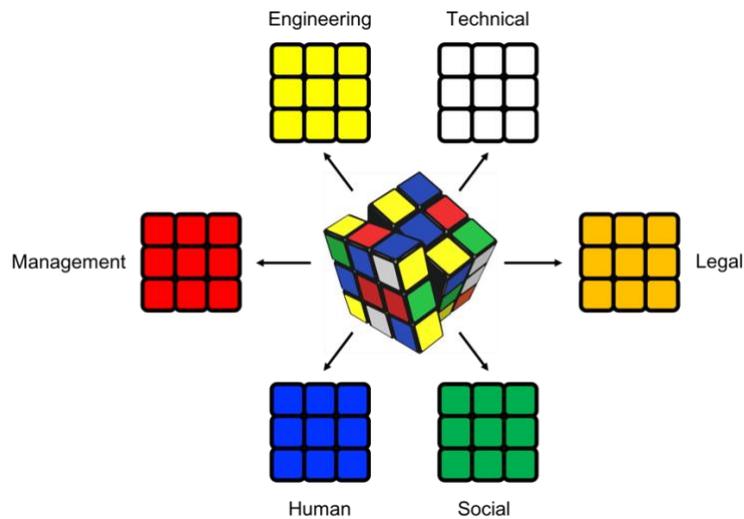

Figure 1.1 Systems Engineering as a Rubrick's Cube

Kossiakoff *et al.* (2011) suggest that the domains of systems engineering can be described like faces on a Rubik's cube with each face of the cube representing a key domain (Figure 1.1). Specifically, they list the following domains as being central to systems engineering:

- **Engineering:** Applies tools to support system design, such as Computer-Aided Design (CAD) tools, modeling and simulation tools, version control tools, etc.
- **Technical:** Incorporates the technical knowledge from the appropriate engineering disciplines, such as electronics, embedded software, propulsion, materials, etc.
- **Management:** Utilizes project management skills, such as scheduling, cost estimation, resource allocation, risk assessment, etc.
- **Legal:** Focuses on issues surrounding contracting, generation of intellectual property, licensing, subcontracting, etc.
- **Human:** Addresses interpersonal dynamics, team building, employment, etc.; could also be interpreted to include human factors engineering; however, this is better placed in the context of one of the technical disciplines.
- **Social:** Assesses and influences the design in ways that make positive impacts on the world around us. Evaluates those impacts on societal, economic, political, and environmental factors.

As one might expect, these "faces" are not independent. They interact in ways similar to the challenges associated with solving the rubric's cube. By turning one face, it can affect the neighboring faces in unanticipated ways. Thus systems engineering needs to apply rigorous, disciplined approaches at the intersection of these domains to ensure an effective engineering process.

## 1.2 Systems Engineering

In general, when one thinks about approaching an engineering design problem, they start by considering the specific engineering discipline central to that design. Such engineering disciplines include electrical, civil, mechanical, computer, industrial, petroleum, chemical, biological, etc. Systems engineering, however, sits above these disciplines since it is focused on how to draw the technical aspects of several of these disciplines together to create a complete, integrated system.

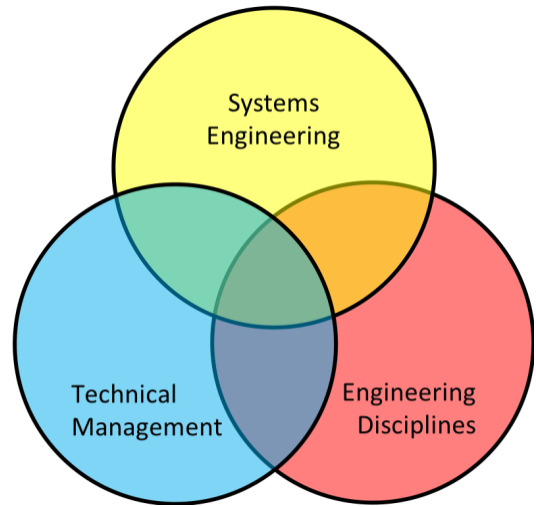

Figure 1.2: Systems Engineering Fields

### 1.2.1 Disciplines in Systems Engineering

To understand systems engineering, we can consider the Venn diagram in Figure 1.2. Here we see that the system design process draws from skills specific to systems engineering as well as the relevant engineering fields and technical management. In addition to these major disciplines, we also draw on the following sub-disciplines, each of which lies at the intersection of these three areas.

- **Modeling and Simulation:** Computer-based models are developed to validate system performance and effectiveness, as well as to determine whether or not specific requirements are being satisfied by the design. Modeling and simulation is usually employed by the engineering disciplines.

- **Operations Research:** Systems design and engineering is a process that involves making decisions that trade, for example, specific performance characteristics with the cost to obtain the desired performance. Operations research is a field devoted to optimal decision-making and includes several areas relevant to system design, including financial engineering, decision analysis, and logistics and supply chain management.

- **Project Management:** Developing a complex system is, by necessity, a team-based process. To be successful, the systems engineering process draws on sound project and technical management to ensure the right resources are available to the team, proper communication occurs among team members, and schedules and deliverables are maintained.

- **Quality Assurance:** Fundamentally, system or product quality depends on system/product health. Quality assurance applies design principles to maximize system quality and incorporates process monitors to ensure that quality is maintained. Technical debt, which is a measure of system quality, is a concept that has arisen in the software engineering discipline that captures the cost of trading strategic decisions in system design for tactical decisions (Allman, 2012). The marketing slogan for FRAM oil filters of "pay me now, or pay me later" is captured formally when estimating technical debt. For example, one of the first decisions in system design that increases technical debt (by increasing life-cycle cost) is to delay addressing system test/system health requirements.

The principal challenge in systems engineering and systems design is balancing the various disciplines. Designing a system that satisfies a vast array of requirements, including functional requirements, reliability and availability requirements, performance requirements, packaging requirements, and system maintenance requirements, requires careful management and utilization of the right technical resources.

### 1.2.2 System Decomposition

One of the key elements in designing a complex system is arriving at an appropriate decomposition of the system into manageable parts. A variety of approaches have been developed to accomplish this. Generally, at the core of such decomposition, is a recognition that systems are often hierarchical in nature. Systems are made up of subsystems, which are made up of sub-subsystems and so forth until we get to the constituent parts. Reversing this, we can think of each of these intermediate steps as building blocks for the next level up in the hierarchy.

Often, when considering how to decompose a system into its constituent building blocks, we need to consider both the physical decomposition as well as the functional decomposition. For example, consider a personal computer (PC) as our system. The PC system is decomposed physically into a monitor, keyboard, and computer, where the computer is then decomposed into a motherboard, a power supply, etc. Things start to get a bit murky in the hierarchy when we consider the different types of cards and integrated circuits in the system. For example, should we regard a GPU that is plugged into the backplane at the same level as the CPU that is plugged directly into the motherboard? Each of them provide processing capabilities, but one is an IC and one is a card.

This is when we run into the issues of functional decomposition. In the same PC, a CPU and a GPU provide similar functional capabilities (albeit differently) so, functionally, they may be considered to be at the same level. However, consider a single dual in-line memory module (DIMM). We may also wish to group several memory modules together by plugging them into multiple slots on the motherboard so as to form a functional unit corresponding to the computer's random access memory (RAM). In this case, we cross the simple physical boundaries of individual DIMM cards and simply combine all of the memory IC's (and supporting circuitry) together.

In modern systems, we also begin to blur the lines between hardware and software. A given function often requires both, so pure functional decomposition breaks down. Similarly, with the increased prevalence in using field programmable gate arrays (FPGA), the distinction between hardware and software blurs. This tends to drop us into the realm of hardware/software co-design. Schaumont (2010) describes the hardware/software co-design space where the intended application for the system being designed defines the design context. We can then think of different levels of complexity for different hardware architectures. For example, ranging from the general purpose reduced instruction set computer (RISC)-based systems that provide maximum flexibility but minimum efficiency to application-specific integrated circuit (ASIC)-based systems that provide minimum flexibility but maximum efficiency, the software is integrated into the respective designs using a variety of programming languages and tools.

Once the appropriate model of system decomposition is determined, the designer must then consider the nature of the interfaces between the various system building blocks. The design of the hardware interfaces must take into account physical constraints, pin layouts, and routing/connectivity with other system components. Software interfaces must consider the nature and type of data being passed between the components. Once again, these two domains become blurred when we are connecting hardware to software. For example, bus systems provide the physical backbone over which signals propagate within a system, and the signals encode the data to be transmitted.

Different approaches exist for handling the hardware/software interface (Schaumont, 2010). The memory-mapped interface uses memory within a microprocessor as the means for communication via load/instructions. The coprocessor interface employs a separate coprocessor port on a microprocessor as the means for using software to control the hardware. Finally, a custom-instruction interface can be used to integrate custom, software-controlled functionality directly into the microprocessor architecture.

As a final concept to be introduced here, system design has also been influenced by the software engineering community by employing object-oriented design principles (Kumar et al., 1994; Bahill and Daniels, 2003; Drabik, 2011). Object-oriented design is based on the idea that a system can be decomposed into logical "objects" that encapsulate behavior within the system (Booch et al., 2007). The interfaces between objects are defined based on methods that implement the behaviors of those objects. Objects are generalized into "classes," which act as abstract data types and can be organized hierarchically using mechanisms such as inheritance (subclasses inherit and possibly specialize properties of superclasses) and polymorphism (the notion that the same interface can be used to interact with objects of different types (Cardelli and Wegner, 1985)).

There exists considerable debate about the appropriateness of the object-oriented design principle when designing hardware, which is somewhat surprising given the fact hardware systems are inherently object-oriented. Object-oriented principles, however, have been demonstrated in software design to lead to more maintainable code as well as better code reusability. With more and more software integration with hardware and hardware/software co-design, the need is becoming greater to employ the same design principles that are so successful in software; therefore, when we discuss software design issues later, the object-oriented approach will tend to stand out.

### 1.2.3 Managing System Complexity

It is a given that modern system design involves designing and assembling the building blocks of what will, ultimately, be highly complex. Issues related to power, behavior, packaging, health management, and support, introduce competing objectives that make the design process difficult. Therefore, it is essential that the design process take steps to manage this complexity from the outset.

Where complexity management begins is with the requirements. Specifying requirements involves imposing constraints on the system design that influence, not only the specific part of the system satisfying those requirements but related parts of the system as well. Requirements interact and define what, ultimately, is feasible. In fact, we can regard the entire system design process as a large multi-objective, constrained optimization problem where we attempt to optimize competing factors (e.g., maximize performance, minimize energy, minimize manufacturing cost, maximize availability, etc.) within the constraints of the requirements (e.g., geo-locate to within one meter, fly at Mach 1.5, communicate with frequency hopping over a set frequency range, provide 100% fault isolation to three or fewer replaceable units, etc.).

Recognizing the role the requirements play, not only in specifying what needs to be done but also in determining the complexity of the resulting system, the task then becomes managing the resulting complexity. Standard approaches employ hierarchical decomposition into appropriate building blocks and careful partitioning between hardware, software, and mixed functional components. This presents us with a challenging modeling problem where, ideally, we would like to encapsulate the systems by their physical or functional boundaries, but interactions across these boundaries will remain that need to be modeled. Additional complications arise when we attempt to mix the results from multiple models. For example, attempting to combine a digital simulator with the results of finite element analysis to determine how G-forces on the structure of an on-board digital signal processor affect the DSP's functionality is difficult.

An added complication arises when we recognize that many modern systems are actually *systems of systems*. Under such conditions, the design must address issues of unexpected emergent behavior. In general, it is impossible to be able to detect all emergent behavior, but by involving a strong team of domains that work together, some emergent behavior can be anticipated and, if deemed negative, prevented. This may require specialized models to address manageable levels of possible system interactions, usually at the interfaces to determine the extent to which encapsulation has succeeded.

## 1.3 A Systems Engineering Process

The field of systems engineering has emerged as the over-arching discipline for designing and managing complex systems. Systems engineering uses a combination of technical and management skills to provide the framework within which system design occurs. In this section, we take a look at elements in a typical systems engineering process. This includes addressing needs for the design itself, as well as the technical management of the process and ultimate realization of the product.

### 1.3.1 System Design

The classical design process follows what is common referred to as a *waterfall* model. This model is depicted in Figure 1.3. In this model, the engineering process starts with requirements analysis and proceeds through successor stages of system design, implementation, test and verification, and maintenance. Unfortunately, this model, while logical, has been demonstrated to not be the best systems engineering approach for managing complexity.

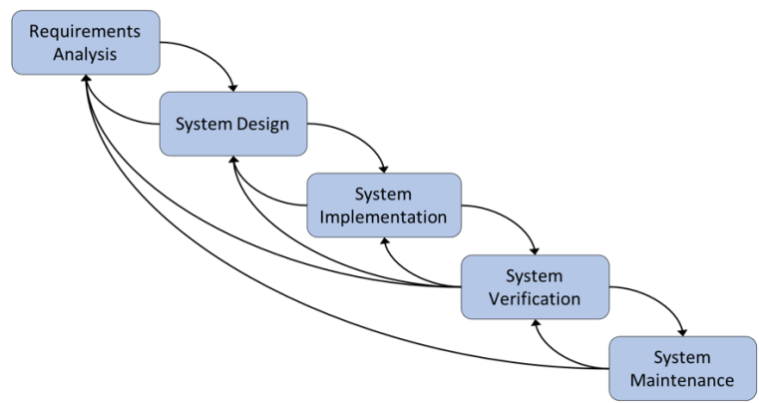

Figure 1.3: Waterfall Model of Systems Engineering

An alternative model has emerged from the software engineering community known as the *spiral* model (Boehm, 1986). The principal idea emerging from this model is one of successive refinement. An initial set of requirements is formulated (referred to as "conceptual" requirements), from which a prototype is built. Each turn through the spiral follows the classic waterfall model; however, at the end of the validate stage, a refinement step begins. With each refinement, stronger requirements are established, and the system continues to evolve until it reaches the final version to be fielded.

Additional systems engineering models have been proposed and even standardized. In EIA 632, a top-down engineering process is defined with a considerable amount of feedback. As described in the standard, "Appropriate processes of [the systems engineering process] are applied recursively and iteratively to define the system products of the system hierarchy from the top down, and then, to implement and transition the system products from the bottom up to the user or customer (EIA, 2003)." Thus the intent of EIA 632 is to provide a capability that combines the traditional top-down design of the waterfall model with the iterative improvement of the spiral model.

Slightly more recent than the EIA standard, the IEEE Software and Systems Engineering Standards Committee in the IEEE Computer Society published a related standard focused on interdisciplinary systems engineering (IEEE, 2005). This standard has been "dual-logoed" with both the International Organization on Standardization (ISO) and the International Electrotechnical Commission (IEC). Similar to the EIA standard, IEEE 1220 combines a waterfall-like process in the midst of iterative refinement. This particular standard also provides detailed processes for each of the main steps that correspond to the classic waterfall process. Also of note is the strong dependence, not only on requirements analysis, decomposition, and allocation, but on conducting trade studies and assessments at each major stage to assess the impacts of the various design decisions.

The last standard we consider is also published by the IEEE and also dual-logoed with ISO and IEC (IEEE, 2015). In this standard, rather than depict a flow-type engineering model, the standard simply identifies what it sees as the major activities in the systems engineering process (Figure 1.4). Key to this approach is that any (or all) of the identified processes can be applied at any stage of development. It does not depend on a particular model of systems engineering, so it works in a waterfall environment, a spiral environment, or any other.

| Agreement Processes | Technical Management Processes | Technical Processes |
|---|---|---|
| • Acquisition<br>• Supply | • Project Planning<br>• Project Assessment and Control<br>• Decision Management<br>• Risk Management<br>• Configuration Management<br>• Information Management<br>• Measurement<br>• Quality Assurance | • Business or Mission Analysis<br>• Stakeholder Needs and Requirements Definition<br>• System Requirements Definition<br>• Architecture Definition<br>• Design Definition<br>• System Analysis<br>• Implementation<br>• Integration<br>• Verification<br>• Transition<br>• Validation<br>• Operation<br>• Maintenance<br>• Disposal |
| **Organizational Project-Enabling Processes**<br>• Life Cycle Model Management<br>• Infrastructure Management<br>• Portfolio Management<br>• Human Resource Management<br>• Quality Management<br>• Knowledge Management | | |

Figure 1.4: IEEE Standard 15288 Systems Engineering Process

### 1.3.2 Technical Management

One of the elements in systems engineering that becomes more obvious in is the role of technical management. According to ISO/IEC/IEEE 15288

> The Technical Management Processes are concerned with managing the resources and assets allocated by organization management and applying them to fulfill the agreements into which the organization or organizations enter. The Technical Management Processes relate to the technical effort of projects, in particular to planning in terms of cost, timescales, and achievements to the checking of actions to help ensure that they comply with plans and performance criteria, and to the identification and selection of corrective actions that recover shortfalls in progress and achievement. They are used to establish and perform technical plans for the project, manage information across the technical team, assess technical progress against the plans for the system products or services, control technical tasks through to completion, and aid in the decision-making process (IEEE, 2015).

The standard identifies eight processes that, generally speaking, capture different tasks of technical management regardless of the model implemented. These tasks include:

- **Project Planning:** This process is focused on ensuring effective and workable plans are created and maintained throughout the design process. Relevant plans include, but are not limited to, development plans, quality plans, test plans, documentation plans, build plans (for software), deployment plans, and maintenance plans.

- **Project Assessment and Control:** Given the plans that emerge from project planning, this task determines whether or not the plans are workable, feasible, and aligned with each other. It is in this context that program reviews and technical progress reviews are conducted.

- **Decision Management:** Throughout the design process, multiple design trade-offs will need to be considered. Through this process, a framework will be provided by which such decisions can be made relative to design objectives and requirements.

- **Risk Management:** Historically, this process is focused on assessing and mitigating the risks associated with the engineering process. In this case, risk is defined very broadly so it can include anything related to managing uncertainty in the engineering process, including impacts on schedule, cost, environmental impact, and safety. This process is distinct from the type of risk analysis that needs to be included within the context of operations and maintenance,

- **Configuration Management:** Critical in any complex engineering activity is ensuring a clear understanding of system configuration and version are maintained. Within the software engineering community, version control is maintained through code repositories such as Subversion (svn) (Pilato et al., 2010) or more recently, Git (Chacon and Straub, 2018). Similar ideas are applied in full system development in terms of maintaining versions of documentation, plans, diagrams and drawings, models, etc. An overarching configuration management process needs to be in place to correlate all of the various versions to the appropriate system configurations.

- **Information Management:** With any systems engineering activity, a set of stakeholders drive the requirements and the development process. As the design unfolds, considerable amounts of data and information are generated. This information, which includes things like reports, plans, designs, test results, and documentation, needs to be managed in a way that is accessible to the stakeholders.

- **Measurement:** The purpose of the measurement process is to capture, analyze, and report the results of the entire design process to assess progress, quality, and overall effectiveness. This information is used to guide the decision-making processes throughout the system lifecycle.

- **Quality Assurance:** Fundamentally, the goal of any engineering activity is to design and implement/manufacture a quality product. Quality plans are produced during the project management process, and quality is assessed through the measurement process. The quality assurance process is focused on ensuring that the plans are implemented correctly and the quality indicators are analyzed to assess the quality of the process and product.

While considered a Technical Process by the standard, rather than a Technical *Management* Process, we assert that Operations and Maintenance Planning falls within the purview of Technical Management. Ideally, the type of management activities performed here are addressed throughout the design cycle, rather than waiting until deployment or post-deployment.

- **Operations:** The operations stage of a product is the period at which the product is performing its intended function. While technically this is not part of the design phase, it is an important part of the lifecycle of the product. From a Technical Management perspective, operations addresses the process by which the product is used, performance is managed, and assessments are made to determine whether or not the product is performing as intended. From a health perspective, operational management is critical, so appropriate plans and procedures need to be put in place for system monitoring during the design phase.

- **Maintenance/Support:** Similar to the operations stage, the maintenance (or more broadly, support) stage of a product is the period in which some action needs to be executed on the product to enable that product to continue to perform its function. Maintenance involves inspections, anomaly detection, fault detection and isolation, repair, and overhaul. During the maintenance phase, the system is not necessarily available for use, unless some sort of redundancy or multi-model operational capability is provided by the product. As with the operations stage, maintenance and support management is critical, necessitating the development of proper plans and procedures during the design phase of the product.

More detail will be given on the Operations and Support aspects of a product in Section 1.4.

### 1.3.3 Product Realization

Once a system has been designed, it is time to realize that system—to turn it into an actual product. Typically, a number of steps are required for such product realization, ranging from implementing the system from the bottom up, integrating the components into the next level in the product hierarchy, and along the way, verifying and validating that the system meets its intended requirements. Once realized, the system then needs to be transitioned into actual use.

One way of depicting the process is described in (NASA, 2007). This process consists of three major phases that comprise five steps. The first phase focuses on design realization where the components of the system are implemented, combined, and integrated. The second phase verifies the product to verify that it functions in accordance with its requirements and validates the product to ensure that it will function as intended in its target environment. Finally, the product is transitioned into the field. More specifically, the NASA Product Realization Process consists of the following:

- **Implementation:** The first step of the product realization process is to implement the product itself. Product implementation may involve manufacturing hardware, coding software, acquiring off-the-shelf components, reusing products that have been created for some other use, incorporating simulation models, or otherwise generating an actual artifact based on the system design. It is customary that, as components are created, they are also tested to determine whether or not that portion of the system functions as intended. Such testing can include unit tests, interconnect tests, functional tests, etc.

- **Integration:** As product components or subassemblies are generated, they need to be assembled to form the next higher level in the product hierarchy (sometimes also referred to as the product breakdown structure, or PBS). As with the implementation phase, integration must include testing to ensure that the components work together as intended and do not yield any unexpected, emergent behavior. Much of the testing that occurs during integration focuses on the interfaces between the components, including both the physical and the functional interfaces.

- **Verification:** The main question to be addressed by verification is whether or not the product was realized properly. More specifically, the purpose of verification is to ensure that the realized product meets the design requirements. One can argue that, no matter how useful a product might be, if the product does not satisfy the design requirements that led to its realization, then the product is not implemented properly. Verification is performed through formal analysis, demonstration, inspection, and testing against acceptance criteria. As part of the verification process, features of the realized system are traced back to the requirements that spawned them to ensure that all of the requirements have been met.

- **Validation:** In contrast to verification, the main question to be addressed by validation is whether or not the correct product was realized. More specifically, it is the goal of validation to ensure that the system functions as intended in its target environment and in accordance with its intended use. It is customary for a complex system to have a "concept of operations" (ConOp) defined to describe its intended use. In software development, such a ConOp often emerges as a part of "use case analysis." As part of the verification process, features of the realized system are traced back to the requirements that spawned them to ensure that only the stated requirements have been realized.

- **Transition/Deployment:** Product transition and deployment is an incremental process. The goal of product transition is to move the product into the environment within which it is intended to be used. Product transition occurs throughout the product lifecycle, including in the overall product realization process. It involves handling the move from design and development to production. In addition, as a component is realized, it is transitioned into its place within the product hierarchy; therefore, transition can be a part of product integration. It includes handling configuration management and version control. The final transition of a full system into the field corresponds to deployment and incorporates supporting products and processes such as packaging, containers, storage, shipping, and receiving.

## 1.4 Operation and Support

Finally, we consider the operation and support (O&S) phase of the system life cycle. Typically, we consider O&S to be outside of the scope of system design and systems engineering; however, given a common lack of focus on designing for system health, we assert that the system design process must plan for how the system's health will be monitored and maintained once in the field. Therefore, it is not realistic to expect to be able to divorce the design process from the O&S process.

### 1.4.1 Design for Testability

Historically, system health has been addressed by designing the system with specific test capabilities. As an example, one of the most common approaches to designing testability into an integrated circuit is through the use of Boundary Scan IEEE (2013). The basic idea is to place a set of flip-flops at the pins around the periphery of an integrated circuit such that signals can be passed in and out, effectively isolating the IC from the rest of the system.

A typical implementation of boundary scan is depicted in Figure 1.5. Here we see a representation of an integrated circuit with the core logic of the circuit highlighted in the middle. Around the periphery of the circuit is added a set of boundary registers/flip flops that are used to capture test inputs and outputs. These registers are connected directly to the pins of the IC and feed into the core circuitry. The registers are also connected in series to permit patterns to be passed into the registers via the Test Data In (TDI) port and read out serially through the Test Data Out (TDO) port. The IC also includes additional logic corresponding to a Test Access Port (TAP) Controller that determines the behavior when placed into test mode vs. operational model. The TAP Controller is managed through a separate Test Clock (TCK) and Test Model Select (TMS).

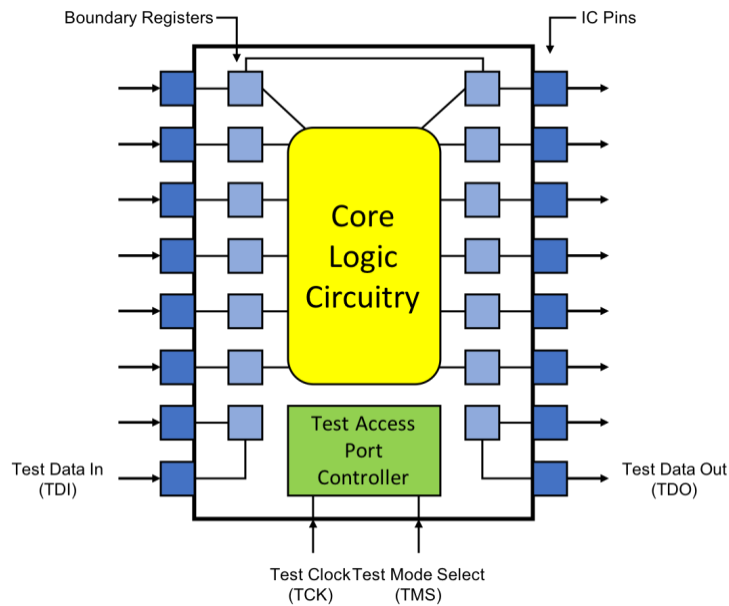

Figure 1.5: Boundary Scan Architecture for an Integrated Circuit

Another approach to designing testability into an electronic system is to incorporate Built-In Test/Built-In Test Equipment, which acts as a separate monitor on the target system to measure performance. BIT/BITE incorporates logic (software and hardware) and circuitry directly into the system to monitor the system for different error conditions. Figure 1.6 gives an example of a notional system with BITE available for doing on-board testing.

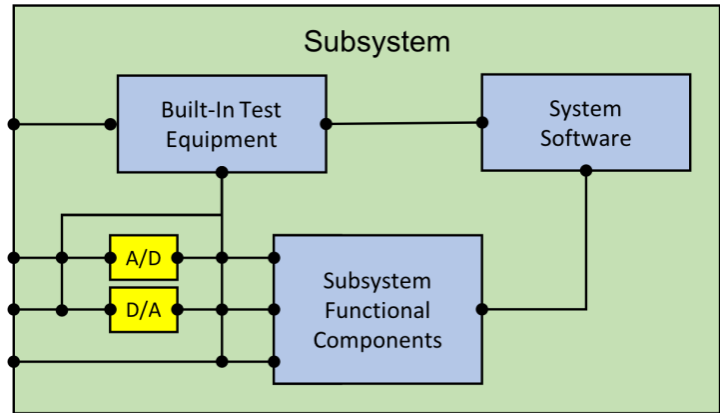

Figure 1.6: Design for Testability with Built In Test Equipment

Four different types of BIT are often employed in a complex system (Butler, 2006). These include Startup BIT (SBIT), Continuous BIT (CBIT), Initiated BIT (IBIT), and Maintenance BIT (MBIT). SBIT is used to ensure the system is ready for use, prior to starting up the application on the unit. CBIT runs continuously in background while the unit is operating and must function in a non-interfering fashion. IBIT can be triggered during operation should unexpected conditions arise, or it can be used when the system is offline to assist in fault isolation. MBIT is run purely offline when the system is in maintenance mode.

A number of methods exist for implementing built-in test, either as part of the normal system circuitry or using separate BITE. Techniques most often applied include things like parity checks, checksums, sequence checks, memory scans for stuck-at faults, and heartbeat checks. These are all tests that check the integrity of the information being passed into and out of the components of the system to verify that this information has not been corrupted in some way. At times, additional methods based on pre-defined test patterns can be stored in read-only memory or non-volatile random access memory that can then be used to test the functionality of the components as well.

### 1.4.2 Maintenance Architectures

Often, a variety of maintenance architectures are employed to provide field support for a system (Simpson and Sheppard, 1994). For complex systems such as aircraft, maintenance usually occurs in various levels, the most frequently employed architecture being a three-level maintenance architecture. For the military, the three levels are often referred to as the "organizational level" (O-level), "intermediate level" (I-level), and "depot level" (D-level). Commercial airlines often employ a similar architecture for maintaining their aircraft. Of note, however, is that many maintenance architectures are being streamlined, either by eliminating a level, merging levels, or bringing the original equipment manufacturer (OEM) into the process. In settings such as the automotive industry, there is often only one level of maintenance (unless we include the automobile owner performing their own maintenance), and personal computers also often employ a single level of maintenance (unless the PC needs to be sent back to the OEM).

The process by which faults are identified so that the system can be repaired is essential to whatever maintenance architecture is employed. Key to understanding such processes is a clear understanding of what we mean by "diagnosis." For our use, we define diagnosis as follows

**Definition 7.** Diagnosis is a functional capability to determine that an equipment malfunction has occurred and to determine the cause of the malfunction to support operator decisions, maintainer decisions, and automated failure responses.

With this definition, we can address four aspects of diagnosis: 1) fault detection, 2) fault localization, 3) fault isolation, and 4) fault prediction. Note that we include fault prediction as part of diagnosis since we claim that prognosis is no more than applying diagnosis in a temporal context. We repeat definitions for the first three from Simpson and Sheppard (1994).

**Definition 8.** Detection is the ability of tests or observations to assert that a failure or malfunction in some system has occurred.

**Definition 9.** Localization is the ability of tests or observations to assert that a fault has been restricted to some subset of possible causes of failure or malfunction.

**Definition 10.** Isolation is the ability of tests or observations to assert that a specific fault is the cause of failure or malfunction. In the context of a specific maintenance level, isolation is localization to a subset of possible causes sufficiently small to lead to the repair of a single replaceable unit at the given maintenance level.

Key to any maintenance process is control of the timing over which maintenance occurs. To that end, the least desirable type of maintenance is what is referred to as "reactive" or "unscheduled" maintenance. As the name suggests, unscheduled maintenance occurs when a system must be taken offline for purposes of performing maintenance or repair as a result of some unexpected event (e.g., a failure) occurring. It should be evident that the disruptive nature of unscheduled maintenance is undesirable and is the result of poor planning in the operation and support of the system.

To prevent unscheduled maintenance, a variety of alternative models have been suggested. One of the most common strategies is to apply scheduled or periodic maintenance. In automobiles, this occurs when we change the oil every 3000 miles and run a general maintenance check perhaps once per year. Commercial airlines perform periodic checks based on number of hours flown (DOT, 2018). For example, the A- check occurs every 400-600 flight hours. This is like the car oil change where filters are replaced, control surfaces are lubricated, and all emergency equipment is inspected. A D-check, on the other hand, involves a complete inspection and overhaul of the aircraft and occurs every 6-10 years. Such inspections can lead to additional maintenance being performed.

### 1.4.3 Condition-Based Maintenance

Traditional approaches to viewing O&S of systems involve considering the normal product lifecycle with interruptions in system availability arising due to periodic inspections, scheduled maintenance, system upgrades, and unscheduled maintenance (failures). Recently, the US Department of Defense has turned to adapting its O&S processes to exist within the context of so-called "condition-based maintenance" (CBM) and has published a DoD instruction implementing a program referred to a CBM+ "as a principal consideration in the selection of maintenance concepts, technologies, and processes for all new weapon systems, equipment, and materiel programs based on readiness requirements, life cycle cost goals, and reliability centered maintenance (RCM)-based functional analysis formulated in a comprehensive reliability and maintainability (R&M) engineering program (DoD 4151.22, 2012)."

CBM+ is defined to be "the application and integration of appropriate processes, technologies, and knowledge-based capabilities to improve the reliability and maintenance effectiveness of DoD systems and components (DoD, 2008)." This program emerged due to the high costs associated with O&S (taking up as much as 65-80% of a system's life cycle cost), where much of the cost can be attributed to system maintenance. Even though CBM+ is defined within the context of DoD systems, the concepts apply to any complex system. Therefore, in the following, we approach O&S from the perspective of CBM+ with a particular focus based on Prognostics and Health Management (PHM).

### 1.4.4 Prognostics and Health Management

As more and more systems move into the realm of CBM and CBM+, we see maintenance decisions being made based on reliability predictions (so-called reliability-centered maintenance, or RCM) and the monitored condition of the system (thus condition-based). RCM and CBM+ have led the way to incorporating prognostics and health management (PHM) into the overall maintenance strategy whereby maintenance is performed in more of an anticipatory, "just in time" sense. A standard published recently by the IEEE focuses on defining a framework for PHM and maps the main activities to the layers of the OSA-CBM framework (Machinery Information Management Open Standards Alliance (MIMOSA), 2017).

Specifically, OSA-CBM defines a layered framework designed to identify the information to be exchanged by the components of a CBM system. IEEE Std 1856 marries OSA-CBM and PHM capabilities into a general system support framework. The OSA-CBM standard consists of a layering of six functional blocks designed to encompass the full range of functionality required for CBM and PHM systems layered from bottom to top as follows:

1. **Data Acquisition (DA):** The DA block gathers raw data from sensors and transducers about the system being monitored.

2. **Data Manipulation (DM):** The DM block is responsible for performing signal processing, signal transformation, and feature extraction on the acquired data.

3. **State Detection (SD):** The SD block compares and analyzes the processed data with expected values or system models to discover abnormalities in the data.

4. **Health Assessment (HA):** The HA block performs fault detection and fault isolation to find the root cause of system abnormalities.

5. **Prognostics Assessment (PA):** The PA block generates predictions about the future health state of the monitored system and the progression of detected abnormalities and faults.

6. **Advisory Generation (AG):** The AG block suggests potential maintenance actions and alternative operating instructions for the duration of the mission that assist in decision-making logistics actions.

IEEE Std 1856 adds two additional layers at the extremes (IEEE, 2017).

7. **Sensors (S):** The S block as the bottom layer corresponds to the physical layer of the CBM/PHM framework where raw sensor data is collected. It also incorporates any other health-related information that might be moving through the system.

8. **Health Management (HM):** The HM block as the top layer takes the information provided from AG to determine how to respond to keep or restore the system to a healthy state. This is the layer that is ultimately responsible for implementing the recommendations from AG.

Given the OSA-CBM framework, the PHM process follows a logical sequence of actions: Sense (system condition) $\Rightarrow$ Acquire (information about the system from the sensors) $\Rightarrow$ Analyze (the acquired information to determine health state) $\Rightarrow$ Advise (on appropriate actions to take based on health state) $\Rightarrow$ Act (to carry out the recommendations and maintain system health). After taking action, the system is placed back in service, and the process continues.

Finally, the O&S process needs to be sensitive to logistics, sparing, and working with system warranties. For our purposes, logistics consists of the processes by which we manage people, goods, and services in support of a system. The objective is to maximize the availability of a system but to do so in an economically efficient manner. To that end, we can consider logistics as consisting of the following stages (Morana, 2018):

1. Identify market needs in terms of quality of service.
2. Design the logistics system in terms of product needs and constraints.
3. Develop the logistics system to meet the requirements associated with product support.
4. Implement the industrial processes necessary to put the logistics system in place.
5. Distribute the necessary items (spares, goods, consumables).
6. Support the product using the implemented system.
7. Refine the logistics process based on feedback of effectiveness, availability, and cost.

## 1.5 Conclusion

As the title of this book indicates, the endeavor we explore here is that of realizing *complex integrated systems*. Fundamentally, this book is about systems engineering but on a large scale. As laid out in this chapter, we seek to bring together and to employ the wide range of methods, tools, and techniques necessary to manage such complex systems, hopefully drawing upon the best such methods and identifying where detailed information can be found to learn more about those methods, tools, and techniques.

This book approaches system design from an *integrated* perspective, where one of the key points of integration involves keeping operations and support of the system at the forefront of the design, development, and implementation process. Historically, a figurative (and sometimes literal) wall was constructed between the design engineers and the support engineers. That approach is untenable, especially in an age of highly complex systems. Therefore, a theme runs throughout this book, reminding the reader of the importance of making O&S an integral part of the life cycle engineering process from the beginning.